\documentclass[sigconf]{acmart}

\setcopyright{none}
\renewcommand\footnotetextcopyrightpermission[1]{}
\acmConference[KDD'26 Workshop]{SciSoc Agents \& LLMs}{August 2026}{Jeju, Korea}
\acmYear{2026}

\usepackage[T1]{fontenc}
\usepackage{booktabs}
\usepackage{multirow}
\usepackage{graphicx}
\usepackage{amsmath}
\usepackage{tikz}
\usepackage{pgfplots}
\pgfplotsset{compat=1.17}
\usetikzlibrary{arrows.meta,backgrounds,fit,calc}
\usepgfplotslibrary{fillbetween}

\begin{document}

\title[Continuous Improvement \& Parallel Autonomous Exploration]{Continuous Improvement and Parallel Autonomous Exploration: An LLM-Agent Framework for Searching Large Solution Spaces}

\author{Dulmini Hettiarachchi}
\affiliation{%
  \institution{Mercari, Inc.}
  \country{Japan}
}
\email{dulminih@mercari.com}

\author{Andre Rusli}
\affiliation{%
  \institution{Mercari, Inc.}
  \country{Japan}
}
\email{andre.rusli@mercari.com}

\author{Julio Christian Young}
\affiliation{%
  \institution{Mercari, Inc.}
  \country{Japan}
}
\email{jc.young@mercari.com}

\author{Sho Akiyama}
\affiliation{%
  \institution{Mercari, Inc.}
  \country{Japan}
}
\email{s-akiyama@mercari.com}

\begin{abstract}
We present a \emph{framework} that gives LLM agents two mechanisms for
searching large solution spaces autonomously.
First, a leaderboard scored on held-out data acts as a reward signal that
drives each agent to refine its solutions over repeated submissions, a loop
that operates even with a single agent.
Second, the framework enables running many agents in parallel, fully autonomously, with
\emph{no human in the loop}: agents independently analyze, survey methods,
implement, self-evaluate, submit, and revise, while a moderator agent handles
only logistics.
Running agents in parallel under the shared reward broadens the explored
region of the solution space rather than refining the single seeded paradigm.
We instantiate the framework on product-to-catalog matching (a core
e-commerce retrieval task with a large, category-structured solution
space), posed as selective prediction with a precision-coverage operating
point.
A single agent refines within its seeded paradigm,
whereas parallel autonomous agents surface qualitatively different solutions.
On this testbed, best qualified coverage ($\geq$95\% P@1 per category) reaches 47.8--57.4\% with a single agent and 62.8--69.4\% with five, against a 33.3\% baseline.
Our contribution is the framework itself: a continuous-improvement reward loop
and a substrate for fully autonomous parallel exploration, backed by
case-study evidence.
\end{abstract}



\maketitle

\section{Introduction}

Large language models (LLMs) have enabled autonomous agents that write code, run experiments, and iterate on results with minimal human intervention~\cite{wang2024survey}.
Many practical optimization problems have a large solution space and often require task-specific optimization strategies.

We present a framework that gives LLM agents two mechanisms for searching such spaces autonomously.
(i) A withheld-test \emph{leaderboard} acts as a reward signal: each submission is scored on data the agent never sees, and the returned score drives the agent to improve its submissions over successive rounds, a loop that operates even with a single agent.
(ii) The framework enables running \emph{many agents in parallel, fully autonomously, with no human in the loop}: agents independently analyze data, survey candidate methods, hypothesize, implement, self-evaluate, submit, and revise; the only shared signal is the leaderboard, and a logistics-only moderator.
Running agents in parallel under the shared reward broadens the explored region of the solution space.

We instantiate the framework on product-to-catalog matching in a large consumer-to-consumer (C2C) marketplace, a core e-commerce retrieval task.
Unlike catalog-fed B2C, C2C listings are individual-seller free text with no structured stock-keeping unit (SKU) feed. Mapping each listing to its catalog SKU therefore underpins search, recommendation, deduplication, and price guidance, making the task both harder and more business-critical.
Given a user listing (title, description, metadata), the system must identify the correct catalog SKU from a catalog of up to 33K SKUs.
We pose this as a \emph{selective prediction} problem~\cite{geifman2017selective}: systems may abstain on low-confidence listings to achieve high precision on the remainder.
The practical goal, automating only high-confidence predictions, creates a precision-coverage tradeoff: maximize coverage subject to $\geq$95\% Precision@1 on every category independently.
Because different categories reward different approaches, no single method dominates; this large, category-structured solution space makes product-to-catalog matching a natural testbed for broad autonomous exploration.

Prior multi-agent LLM work has focused on collaborative or communicative architectures: MetaGPT~\cite{hong2024metagpt} and ChatDev~\cite{qian2024chatdev} coordinate agents through structured workflows or chains, while CAMEL~\cite{li2023camel} uses role-playing cooperation.
We instead study agents that run independently and in parallel under a shared leaderboard reward, sharing only scores and high-level approach descriptions.
While MLAgentBench~\cite{huang2024mlagentbench} evaluates individual agent capability on ML tasks, our framework examines how a reward-driven loop plus parallel autonomous exploration searches a large space.

Our key contributions:
\begin{enumerate}
    \item \textbf{Framework}: we define a continuous-improvement reward loop (a withheld-test leaderboard that scores each submission and drives every agent to iteratively improve its solutions) plus a substrate for \emph{fully autonomous parallel exploration} of large search spaces.
    \item \textbf{Autonomy}: we show the framework runs end-to-end with \emph{no human in the loop} (no human prescribes technical approaches; the moderator is automated and handles only logistics), and that parallel agents under the shared reward broaden exploration beyond a single agent's seeded paradigm.
    \item \textbf{Case-study evidence}: on product-to-catalog matching, we report a finding that replicates across runs and is independent of agent count: a seeded single agent iterates narrowly within its paradigm while parallel autonomous agents escape it; the coverage figures are descriptive ranges, not a causal scaling law.
\end{enumerate}

\section{Related Work}

\paragraph{LLM Agents for Scientific Discovery.}
Autonomous LLM agents drive scientific discovery (The AI Scientist~\cite{lu2024aiscientist}, ChemCrow~\cite{bran2024chemcrow}, Coscientist~\cite{boiko2023coscientist}) and mathematical search (FunSearch~\cite{romera2024funsearch}), and are benchmarked individually on ML and software tasks (MLAgentBench~\cite{huang2024mlagentbench}, SWE-bench~\cite{jimenez2024swebench}).
We complement these by studying how parallel autonomous agents collectively explore an optimization landscape under a shared reward.

\paragraph{Multi-Agent LLM Frameworks.}
AutoGen~\cite{wu2023autogen} provides conversational multi-agent coordination.
MetaGPT~\cite{hong2024metagpt} encodes SOPs into role-based pipelines, ChatDev~\cite{qian2024chatdev} coordinates agents through chat chains, and AgentVerse~\cite{chen2024agentverse} studies emergent group behaviors.
Debate-based methods~\cite{du2023debate} improve factuality through argumentation.
These are primarily \emph{collaborative} designs.
CompeteAI~\cite{zhao2024competeai} studies competition among LLM agents in a social-simulation setting; we instead use a shared leaderboard as a reward signal for parallel agents' autonomous \emph{algorithmic} exploration of a technical search space.

\paragraph{Iterative Agent Optimization.}
ReAct~\cite{yao2023react} interleaves reasoning with tool use, and Reflexion~\cite{shinn2023reflexion} introduces verbal reinforcement learning through self-reflection, directly analogous to our iterative submission-feedback loop.
OPRO~\cite{yang2024opro} treats LLMs as optimizers.
Agent-K~\cite{grosnit2024agentk}, which uses the Kaggle leaderboard as a reward signal for iterative improvement, is the closest single-agent analogue to our first mechanism; AutoKaggle~\cite{li2024autokaggle} similarly optimizes single-pipeline ML workflows, and MLE-bench~\cite{chan2024mlebench} benchmarks such agents. We extend this withheld-test reward loop to parallel agents, where solution diversity emerges from a shared reward rather than an explicit diversity-preservation operator.
CoMind~\cite{li2025comind} is the closest multi-agent analogue; key differences are local training-proxy rewards (versus withheld-test scores), a shared community pool of code notebooks and discussions (versus scores and approach descriptions only), and a specialized-role pipeline (Coordinator, Analyzer, Idea Proposer, Evaluator) rather than independent generalist agents each running the full research loop.
This contrasts with quality-diversity search~\cite{mouret2015mapelites} and FunSearch's~\cite{romera2024funsearch} island populations.

\paragraph{Product Matching and E-Commerce Retrieval.}
DeepMatcher~\cite{mudgal2018deepmatcher} established deep learning baselines for entity matching; Ditto~\cite{li2021ditto} showed fine-tuned pre-trained LMs can outperform prior entity-matching systems.
Dense passage retrieval~\cite{karpukhin2020dpr} popularized the dense bi-encoder retrieval paradigm used by models like BGE-M3~\cite{bge-m3}, with FAISS~\cite{johnson2019faiss} providing nearest-neighbor infrastructure.
Selective prediction~\cite{geifman2017selective} underpins our precision-coverage tradeoff.

\section{Method}

\subsection{Task Definition}

Given a user listing $q = (t_q, d_q, m_q, c_q)$ with title, description, metadata, and category, retrieve the correct catalog SKU $s^* \in \mathcal{S}_{c_q}$, where $\mathcal{S}_{c_q}$ is the catalog for category $c_q$ and each SKU $s = (t_s, d_s, m_s, c_s)$ has the same field structure.
Each prediction includes a confidence score; listings below a threshold are excluded, trading coverage for precision.
The evaluation metrics are:
\begin{itemize}
    \item \textbf{Precision@1} (P@1): Fraction of \emph{covered} predictions where the top-1 retrieved SKU is correct.
    \item \textbf{Coverage}: Fraction of listings for which a prediction is made.
\end{itemize}
\textbf{Qualification} requires $\geq$95\% P@1 on \emph{every} category independently.
Among qualified submissions, the ranking objective is maximum average coverage.
This setup reflects a practical operating point: automate high-confidence matches while routing uncertain listings for manual review.

We evaluate on three categories from a large Japanese C2C marketplace (Table~\ref{tab:data}).
Agents develop methods using the training split (with ground truth for self-evaluation), then submit predictions on the test split, which is evaluated by the leaderboard against withheld ground truth.

\begin{table}[h]
\centering
\small
\begin{tabular}{lcc}
\toprule
Category & Base P@1 & Base Cov. \\
\midrule
Smartphone & 100.0\% & 1.3\% \\
Trading Card (Others) & 94.5\% & 58.6\% \\
Trading Card (Pokemon) & 97.3\% & 39.9\% \\
\midrule
\textbf{Average} & \textbf{97.3\%} & \textbf{33.3\%} \\
\bottomrule
\end{tabular}
\caption{Dataset statistics and BGE-M3 baseline (test split). Thresholds tuned on train for 95\% P@1 per category, then applied to test; the baseline is \emph{not} qualified (Trading Card (Others) $<$95\% P@1).}
\label{tab:data}
\end{table}

\subsection{Framework Infrastructure}
\label{sec:infrastructure}

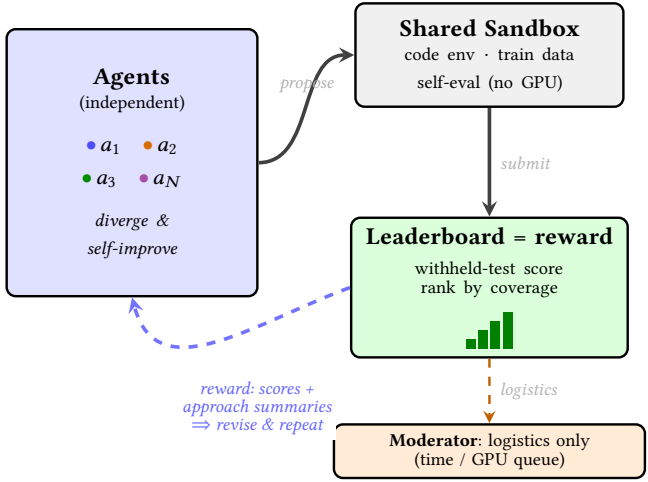
\begin{figure}[t]
\centering
\resizebox{\columnwidth}{!}{%
\begin{tikzpicture}[
  >=stealth, thick, font=\small,
  agent/.style={draw, rounded corners=3pt, fill=blue!12, align=center},
  box/.style={draw, rounded corners=3pt, fill=gray!12, align=center, font=\small},
  lb/.style={draw, rounded corners=3pt, fill=green!14, align=center, font=\small},
  mod/.style={draw, rounded corners=3pt, fill=orange!16, align=center,
    font=\scriptsize},
  flow/.style={->, very thick, black!75},
  fb/.style={->, dashed, very thick, blue!55},
  stage/.style={font=\scriptsize\itshape, text=gray!60},
]

\node[agent, text width=2.7cm, minimum height=3.1cm] (ag) at (0,2.5)
 {\textbf{Agents}\\[1pt]\scriptsize (independent)\\[6pt]
  \footnotesize
  \textcolor{blue!70}{$\bullet$}\,$a_1$\quad
  \textcolor{orange!85!black}{$\bullet$}\,$a_2$\\[3pt]
  \textcolor{green!55!black}{$\bullet$}\,$a_3$\quad
  \textcolor{violet!70}{$\bullet$}\,$a_N$\\[7pt]
  \scriptsize\textit{diverge \&}\\[-1pt]\scriptsize\textit{self-improve}};

\node[box, text width=2.9cm, minimum height=1.2cm] (sb) at (4.15,3.75)
 {\textbf{Shared Sandbox}\\[1pt]\scriptsize code env $\cdot$ train data\\[-1pt]
  \scriptsize self-eval (no GPU)};

\node[lb, text width=3.0cm, minimum height=1.55cm] (lb) at (4.15,1.05)
 {\textbf{Leaderboard = reward}\\[3pt]
  \scriptsize withheld-test score\\[-1pt]\scriptsize rank by coverage\\[4pt]
  {\color{green!50!black}\rule{3pt}{3pt}\,\rule{3pt}{6pt}\,%
   \rule{3pt}{9pt}\,\rule{3pt}{12pt}}};

\node[mod, text width=3.4cm] (md) at (4.15,-0.85)
 {\textbf{Moderator}: logistics only\\[-1pt]
  (time / GPU queue)};

\draw[flow] (ag.east) to[out=0,in=180]
  node[stage, above, pos=0.5, yshift=2pt]{propose} (sb.west);
\draw[flow] (sb.south) -- node[stage, right]{submit} (lb.north);
\draw[fb] (lb.west) to[out=200,in=-75,looseness=1.15] (ag.south);
\node[stage, text=blue!60, align=center, fill=white, inner sep=1.5pt]
  at (1.45,-0.35) {reward: scores +\\[-1pt]approach summaries\\[-1pt]
  $\Rightarrow$ revise \& repeat};

\draw[->, dashed, orange!75!black] (lb.south) --
  node[stage, right, pos=0.5]{logistics} (md.north);

\end{tikzpicture}%
}
\caption{The framework as a continuous, autonomous loop: agents \emph{propose} in a shared sandbox, \emph{submit} to a leaderboard scored on withheld test data, and \emph{revise} from the returned reward (scores and approach summaries, never code). Parallel agents broaden the explored space ($a_1\dots a_N$); no human is in the loop and the automated moderator handles only logistics.}
\label{fig:framework}
\end{figure}

We built a reusable platform (Figure~\ref{fig:framework}) with the following components:

\paragraph{Automated Leaderboard.}
A \texttt{LeaderboardAPI} evaluates submissions against withheld test ground truth that agents never access.
Submissions must include all three categories with per-category confidence thresholds.
The API computes P@1 and coverage per category, checks qualification ($\geq$95\% P@1 on every category), ranks qualified submissions by average coverage, and appends all results to a persistent JSON ledger.
Agents can view the full leaderboard including scores and approach descriptions from all submissions.

\paragraph{Train/Test Separation and Integrity.}
Training data includes ground truth labels for self-evaluation with a lightweight script; test listings contain no ground truth, so agents must submit to the leaderboard for scoring.
Withheld test labels live in an organizer-only location with restricted file permissions. A strict organizer/shared/participant ownership boundary and an automated pre-run leak check \emph{enforce} the separation rather than relying on convention.
Each agent is further confined to its own working directory plus read-only shared resources, so attempts to reuse another agent's artifacts (which we did observe) are structurally prevented rather than merely discouraged.
Each run follows a fixed protocol (setup with automated leak verification, the autonomous run scored on the withheld split by the leaderboard, and an organizer review confirming final standings), so the procedure is reproducible across sessions.

\paragraph{GPU Resource Management.}
Agents share one GPU node; naive concurrent use crashed it in early runs, motivating a two-level lock with a 30-minute \texttt{CLAIM}/\texttt{RELEASE GPU} protocol. GPU access is enforced programmatically by requiring agents to acquire the lock before invoking embedding or model code, and a system-level watchdog terminates unauthorized GPU processes.

\paragraph{Autonomy.}
An append-only chat log is the sole communication channel. \emph{No human is in the loop} beyond a single launch command that specifies agent count and time budget (e.g., ``Run the competition with 5 agents for 5 hours.'').
The moderator is itself an automated agent scoped to logistics only (time warnings, leaderboard updates, queue management); keeping it from drifting into technical guidance was an explicit design requirement. Agents devise and run the full research loop themselves.
Continuous operation is enforced rather than left to agent discretion: each agent writes periodic liveness heartbeats and posts progress to the chat on a fixed cadence, and (instead of terminating after a submission) keeps analyzing, revising, and resubmitting. The moderator emits scheduled warnings (e.g., 30 and 5 minutes before deadline) and forcibly terminates remaining agent processes at the budget deadline.

\subsection{Agent Setup}

Each agent is powered by Claude Sonnet~3.5 (Anthropic) with access to a Python environment including embedding models (BGE-M3~\cite{bge-m3}, Qwen3-Embedding), FAISS~\cite{johnson2019faiss} indexing, and TF-IDF libraries.
Agents receive a task briefing specifying the task, data format, submission protocol, GPU coordination rules, 5-hour time budget, and the BGE-M3 baseline as a starting reference.
Each follows an iterative research loop: analyze data/errors $\rightarrow$ survey candidate methods $\rightarrow$ hypothesize $\rightarrow$ implement $\rightarrow$ self-evaluate on train $\rightarrow$ submit test predictions $\rightarrow$ observe leaderboard $\rightarrow$ revise.

\section{Experiments}

\subsection{Setup}

We contrast two configurations of the framework on identical infrastructure (same tools, same fixed 5-hour per-session time budget with no submission cap), each run three times:

\paragraph{Single agent.}
One agent iterates alone.

\paragraph{Parallel (5 agents).}
Five agents run simultaneously and fully autonomously on a shared GPU.
All coordination is via the append-only chat log; agents see leaderboard scores and approach descriptions but not each other's code.

\paragraph{Baseline.}
BGE-M3 embeddings with cosine similarity and per-category thresholds tuned on train for 95\% P@1, applied to the test split (Table~\ref{tab:data}); not qualified.

\subsection{Results}

\paragraph{Continuous Improvement.}
In both configurations the leaderboard reward drove sustained self-improvement: across runs, best qualified coverage rose from the 33.3\% baseline to 47.8--57.4\% with a single agent and 62.8--69.4\% with five (Table~\ref{tab:main}).
The loop operates even with a single agent, which improved substantially over the baseline through iterative resubmission alone.

\paragraph{Parallel Autonomous Exploration.}
Across all three single-agent runs (60, 124, and 81 submissions) the agent stayed within its seeded embedding paradigm, iterating on BGE-M3 variants and never adopting a non-embedding method; in one run it prematurely judged the hardest category unsolvable and stopped early. In all three 5-agent runs, agents independently reached qualitatively different methods (TF-IDF re-ranking, string-similarity matching) within minutes. The gain concentrates in the hardest category, smartphone (single agents 0.8--1.9\% coverage vs.\ 24.8--41.1\% parallel); on the Trading Card categories a single agent with uncontested GPU is competitive with the parallel runs.
Each configuration was run three times and we report ranges; these illustrate broadened exploration, not a causal effect of agent count, which stays confounded with total compute, GPU contention, and shared leaderboard visibility.

\begin{table}[h]
\centering
\small
\begin{tabular}{lccc}
\toprule
& Baseline & Single agent & Parallel (5) \\
\midrule
Best Avg Cov. & 33.3\% & 47.8 / 57.4 / 49.9 & \textbf{62.8 / 69.4 / 68.2} \\
Best Avg P@1 & 97.3\% & 96.7 / 96.7 / 95.1 & 95.1 / 95.1 / 95.1 \\
Qualified? & No & Yes / Yes / Yes & Yes / Yes / Yes \\
\midrule
Submissions & --- & 60 / 124 / 81 & 274 / 550 / 301 \\
Qualified sub. & --- & 22 / 44 / 46 & 105 / 238 / 103 \\
\bottomrule
\end{tabular}
\caption{Main results: best qualified coverage/P@1 per configuration vs.\ the unqualified 33.3\% baseline; three runs per configuration (run1 / run2 / run3). Descriptive: agent count co-varies with total compute, simultaneous workers, GPU contention, and shared leaderboard visibility; not a causal scaling law.}
\label{tab:main}
\end{table}

\paragraph{Approach Diversity (one representative 5-agent run).}
The five agents ($\alpha$--$\epsilon$ = researcher1--5) all started from BGE-M3 embeddings but diverged: score-gap confidence calibration ($\alpha$, best 50.7\% qualified coverage); BGE-M3 with TF-IDF rescoring ($\beta$, 55.3\%); card-number re-ranking with pure string (SequenceMatcher) matching ($\gamma$, 65.6\%); string-matching plus word/char TF-IDF fusion ($\delta$, 69.4\%, the run's best); and string-similarity re-ranking ($\epsilon$, 63.6\%). All five qualified; the non-embedding string/TF-IDF methods ($\gamma$, $\delta$) drove the top results.

\begin{figure}[t]
\centering
\begin{tikzpicture}
\begin{axis}[
  width=\columnwidth,
  height=4.5cm,
  xlabel={Time (minutes from start)},
  ylabel={Best Qualified Coverage (\%)},
  xmin=0, xmax=370,
  ymin=0, ymax=75,
  xtick={0,60,120,180,240,300,360},
  ytick={0,10,20,30,40,50,60,70},
  tick label style={font=\scriptsize},
  label style={font=\scriptsize},
  grid=major,
  grid style={dashed, gray!25},
  legend style={font=\tiny, cells={anchor=west},
    at={(0.02,0.98)}, anchor=north west,
    inner sep=2pt, column sep=2pt},
  clip=false,
]

\draw[dashed, red!50, thick]
  (axis cs:0,33.3) -- (axis cs:370,33.3)
  node[right, font=\tiny, text=red!50] {baseline};

\draw[dashed, blue!50, thick]
  (axis cs:0,57.4) -- (axis cs:370,57.4)
  node[right, font=\tiny, text=blue!50] {1-agent (57.4)};

\addplot[
  black, very thick,
  const plot mark right,
] coordinates {
  (0,0) (142,27.5) (143,34.8) (162,40.0) (171,50.8)
  (185,60.1) (191,61.1) (207,63.8) (218,65.6)
  (323,65.7) (344,68.8) (357,69.4) (362,69.4)
};
\addlegendentry{5-agent best}

\addplot[
  only marks, mark=triangle*, mark size=2.5pt,
  color=green!60!black,
] coordinates {
  (142,27.5) (344,68.8) (357,69.4)
};
\addlegendentry{$\delta$ (string+TF-IDF)}

\addplot[
  only marks, mark=square*, mark size=2.2pt,
  color=orange!80!black,
] coordinates {
  (191,61.1) (207,63.8) (218,65.6)
};
\addlegendentry{$\gamma$ (card+string)}

\draw[-stealth, thin, gray!70]
  (axis cs:142, 36) -- (axis cs:142, 28.5);
\node[font=\tiny, anchor=south]
  at (axis cs:142, 36.5) {first qualified};

\draw[-stealth, thin, gray!70]
  (axis cs:344, 62.5) -- (axis cs:344, 68);
\node[font=\tiny, anchor=north]
  at (axis cs:344, 62) {string + word-TF-IDF};

\end{axis}
\end{tikzpicture}
\caption{Best qualified coverage over time for one representative 5-agent run (run~2). Dashed lines: unqualified baseline (33.3\%) and best single-agent (57.4\%).}
\label{fig:evolution}
\end{figure}
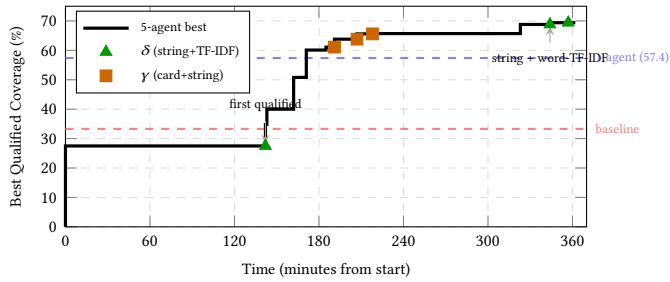

\paragraph{Leaderboard Dynamics.}
In this run the first qualified submission appeared only after extended exploration ($\delta$, BGE-M3, 27.5\%). Coverage climbed through threshold tuning and card-number re-ranking to 65.6\%, then plateaued; the winning push came late, when $\delta$ added string-matching plus word-TF-IDF fusion and reached 69.4\% (Figure~\ref{fig:evolution}).

\section{Discussion}
\label{sec:discussion}

\paragraph{Depth vs.\ Breadth.}
The two mechanisms produce complementary behaviors.
A single agent iterates deeply but narrowly: it refines BGE-M3 variants and thresholds at high precision (avg P@1 95.1--96.7\%), and in one run prematurely concluded the hardest category unsolvable and stopped with budget remaining.
Parallel agents instead widened the search, exploring qualitatively different (non-embedding) strategies rather than deeply refining one, at the cost of lower per-agent precision (95.1\%) and heavy GPU contention.
We read this as a depth-for-breadth trade and do not isolate the cause; the broader exploration nonetheless surfaced string-matching and TF-IDF methods the single agents never considered.

\paragraph{Narrow Exploration vs.\ Escape.}
Seeded with a BGE-M3 baseline, the single agent treated the embedding paradigm as the entire search space: across all three single-agent runs (60, 124, and 81 submissions) it iterated on BGE-M3 variants, score calibration, and thresholds and never instantiated the standard lexical/string baseline a practitioner would try first, in one run prematurely judging the hardest category unsolvable and stopping early.
In all three 5-agent runs, agents reached non-embedding methods (TF-IDF, string-similarity matching) within the session.
That strong lexical/string methods rival dense embeddings on near-identical-variant matching is well known and \emph{not} our claim; the point is behavioral: a single seeded explorer iterates narrowly and prematurely converges, while parallel autonomous exploration escapes it.
Because all runs use one LLM, seed, and prompt, we cannot separate this from a model-specific prior; a no-seed control is the decisive next experiment.

\paragraph{Information Sharing and Integrity.}
Agents could view others' scores and approach summaries but not their code or artifacts: the catalog index is shared infrastructure provided to all, but reusing \emph{another agent's} pre-built indexes or trained models was prohibited.
The reward channel carries both scores and brief approach descriptions (Figure~\ref{fig:framework}), so parallel agents share strategies at a high level while their implementations stay isolated.
This visibility is intended to support differentiation across agents' directions, broadening the explored region of the solution space.

\paragraph{Implications.}
The framework targets applied research: the reward-driven, human-free loop can speed the exploration and improvement of product-to-catalog matching beyond manual iteration and, being task-agnostic, can likewise advance applied research on other tasks in C2C marketplace systems such as search ranking, recommendation, and listing-quality checks.

\paragraph{Limitations.}
Our central qualitative finding (a single seeded agent iterating narrowly and prematurely converging while parallel agents escape) replicates across runs and is independent of agent count, so the limitations below do not affect it.
We make no causal or agent-count claim: agent count co-varies with total compute, simultaneous workers, GPU contention, and shared leaderboard visibility, and the single-vs-parallel coverage figures (Table~\ref{tab:main}) are descriptive, not evidence of a scaling effect. We ran each configuration three times and report ranges. The decisive control for our central finding is a no-seed condition, separating the seeded baseline from a model-specific prior.
All agents use the same LLM (Claude Sonnet~3.5), and the task is bounded (automated metrics, fast iteration); longer-horizon settings with ambiguous success criteria remain open.
We instantiated the framework on a single task (product-to-catalog matching); while it is task-agnostic by design, cross-task generalization to other IR and ML problems remains to be demonstrated empirically.
Agents write and submit code autonomously; individual submissions were not human-reviewed for safety. The infrastructure boundaries described in \S\ref{sec:infrastructure} (per-agent working directories, restricted-permission test labels, the GPU lock) prevent cross-agent damage and integrity violations, but do not audit the LLM-generated code itself.

\section{Conclusion}

We presented a framework that gives LLM agents two mechanisms for autonomously searching large solution spaces: a withheld-test leaderboard that acts as a reward signal driving iterative improvement of submitted solutions, and a substrate for running many agents in parallel with \emph{no human in the loop}.
We instantiated it on product-to-catalog matching, where the central qualitative pattern held across all six runs: the seeded single agent iterated within its embedding paradigm, while parallel autonomous agents reached non-embedding methods (TF-IDF, string-similarity matching) within the session.
The decisive next test is a no-seed control, separating the seeded baseline from the model's own prior; whether these dynamics generalize to other bounded optimization tasks is a further open question.

\bibliographystyle{ACM-Reference-Format}
\bibliography{references}

@article{wang2024survey,
  title={A Survey on Large Language Model based Autonomous Agents},
  author={Wang, Lei and Ma, Chen and Feng, Xueyang and others},
  journal={Frontiers of Computer Science},
  volume={18},
  number={6},
  pages={186345},
  year={2024},
  publisher={Springer}
}

@article{lu2024aiscientist,
  title={The AI Scientist: Towards Fully Automated Open-Ended Scientific Discovery},
  author={Lu, Chris and Lu, Cong and Lange, Robert Tjarko and Foerster, Jakob and Clune, Jeff and Ha, David},
  journal={arXiv preprint arXiv:2408.06292},
  year={2024}
}

@inproceedings{wu2023autogen,
  title={AutoGen: Enabling Next-Gen LLM Applications via Multi-Agent Conversation},
  author={Wu, Qingyun and Bansal, Gagan and Zhang, Jieyu and others},
  booktitle={Proceedings of the First Conference on Language Modeling (COLM)},
  year={2024}
}

@inproceedings{du2023debate,
  title={Improving Factuality and Reasoning in Language Models through Multiagent Debate},
  author={Du, Yilun and Li, Shuang and Torralba, Antonio and Tenenbaum, Joshua B. and Mordatch, Igor},
  booktitle={Proceedings of the 41st International Conference on Machine Learning},
  pages={11733--11763},
  year={2024},
  publisher={PMLR}
}

@inproceedings{bge-m3,
  title={{M3}-Embedding: Multi-Linguality, Multi-Functionality, Multi-Granularity Text Embeddings Through Self-Knowledge Distillation},
  author={Chen, Jianlyu and Xiao, Shitao and Zhang, Peitian and Luo, Kun and Lian, Defu and Liu, Zheng},
  booktitle={Findings of the Association for Computational Linguistics: ACL 2024},
  pages={2318--2335},
  year={2024},
  publisher={Association for Computational Linguistics}
}

@inproceedings{hong2024metagpt,
  title={{MetaGPT}: Meta Programming for A Multi-Agent Collaborative Framework},
  author={Hong, Sirui and Zhuge, Mingchen and Chen, Jiaqi and Zheng, Xiawu and Cheng, Yuheng and Zhang, Ceyao and Wang, Jinlin and Wang, Zili and Yau, Steven Ka Shing and Lin, Zijuan and others},
  booktitle={The Twelfth International Conference on Learning Representations},
  year={2024}
}

@inproceedings{li2023camel,
  title={{CAMEL}: Communicative Agents for ``Mind'' Exploration of Large Language Model Society},
  author={Li, Guohao and Hammoud, Hasan Abed Al Kader and Itani, Hani and Khizbullin, Dmitrii and Ghanem, Bernard},
  booktitle={Advances in Neural Information Processing Systems},
  volume={36},
  year={2023}
}

@inproceedings{qian2024chatdev,
  title={{ChatDev}: Communicative Agents for Software Development},
  author={Qian, Chen and Liu, Wei and Liu, Hongzhang and Chen, Nuo and Dang, Yufan and Li, Jiahao and Yang, Cheng and Chen, Weize and Su, Yusheng and Cong, Xin and others},
  booktitle={Proceedings of the 62nd Annual Meeting of the Association for Computational Linguistics},
  pages={15174--15186},
  year={2024}
}

@inproceedings{chen2024agentverse,
  title={{AgentVerse}: Facilitating Multi-Agent Collaboration and Exploring Emergent Behaviors},
  author={Chen, Weize and Su, Yusheng and Zuo, Jingwei and Yang, Cheng and Yuan, Chenfei and Chan, Chi-Min and Yu, Heyang and Lu, Yaxi and Hung, Yi-Hsin and Qian, Chen and others},
  booktitle={The Twelfth International Conference on Learning Representations},
  year={2024}
}

@article{bran2024chemcrow,
  title={{ChemCrow}: Augmenting large language models with chemistry tools},
  author={Bran, Andres M and Cox, Sam and Schilter, Oliver and Baldassari, Carlo and White, Andrew D and Schwaller, Philippe},
  journal={Nature Machine Intelligence},
  volume={6},
  number={5},
  pages={525--535},
  year={2024}
}

@article{boiko2023coscientist,
  title={Autonomous chemical research with large language models},
  author={Boiko, Daniil A and MacKnight, Robert and Kline, Ben and Gomes, Gabe},
  journal={Nature},
  volume={624},
  number={7992},
  pages={570--578},
  year={2023}
}

@article{romera2024funsearch,
  title={Mathematical discoveries from program search with large language models},
  author={Romera-Paredes, Bernardino and Barekatain, Mohammadamin and Novikov, Alexander and Balog, Matej and Kumar, M Pawan and Dupont, Emilien and Ruiz, Francisco J R and Ellenberg, Jordan S and Wang, Pengming and Fawzi, Omar and others},
  journal={Nature},
  volume={625},
  number={7995},
  pages={468--475},
  year={2024}
}

@inproceedings{jimenez2024swebench,
  title={{SWE}-bench: Can Language Models Resolve Real-World {GitHub} Issues?},
  author={Jimenez, Carlos E and Yang, John and Wettig, Alexander and Yao, Shunyu and Pei, Kexin and Press, Ofir and Narasimhan, Karthik},
  booktitle={The Twelfth International Conference on Learning Representations},
  year={2024}
}

@inproceedings{huang2024mlagentbench,
  title={{MLAgentBench}: Evaluating Language Agents on Machine Learning Experimentation},
  author={Huang, Qian and Vora, Jian and Liang, Percy and Leskovec, Jure},
  booktitle={Proceedings of the 41st International Conference on Machine Learning},
  pages={20271--20309},
  year={2024},
  publisher={PMLR}
}

@inproceedings{yao2023react,
  title={{ReAct}: Synergizing Reasoning and Acting in Language Models},
  author={Yao, Shunyu and Zhao, Jeffrey and Yu, Dian and Du, Nan and Shafran, Izhak and Narasimhan, Karthik and Cao, Yuan},
  booktitle={The Eleventh International Conference on Learning Representations},
  year={2023}
}

@inproceedings{shinn2023reflexion,
  title={Reflexion: Language Agents with Verbal Reinforcement Learning},
  author={Shinn, Noah and Cassano, Federico and Berman, Edward and Gopinath, Ashwin and Narasimhan, Karthik and Yao, Shunyu},
  booktitle={Advances in Neural Information Processing Systems},
  volume={36},
  year={2023}
}

@inproceedings{yang2024opro,
  title={Large Language Models as Optimizers},
  author={Yang, Chengrun and Wang, Xuezhi and Lu, Yifeng and Liu, Hanxiao and Le, Quoc V and Zhou, Denny and Chen, Xinyun},
  booktitle={The Twelfth International Conference on Learning Representations},
  year={2024}
}

@article{li2024autokaggle,
  title={AutoKaggle: A Multi-Agent Framework for Autonomous Data Science Competitions},
  author={Li, Ziming and Zang, Qianbo and Ma, David and Guo, Jiawei and Zheng, Tuney and Liu, Minghao and Niu, Xinyao and Wang, Yue and Yang, Jian and Liu, Jiaheng and Zhong, Wanjun and Zhou, Wangchunshu and Huang, Wenhao and Zhang, Ge},
  journal={arXiv preprint arXiv:2410.20424},
  year={2024}
}

@article{grosnit2024agentk,
  title={Large Language Models Orchestrating Structured Reasoning Achieve Kaggle Grandmaster Level},
  author={Grosnit, Antoine and Maraval, Alexandre and Doran, James and Tutunov, Rasul and Wang, Wenlong and Bou Ammar, Haitham and others},
  journal={arXiv preprint arXiv:2411.03562},
  year={2024}
}

@inproceedings{chan2024mlebench,
  title={{MLE}-bench: Evaluating Machine Learning Agents on Machine Learning Engineering},
  author={Chan, Jun Shern and Chowdhury, Neil and Jaffe, Oliver and Aung, James and Sherburn, Dane and Mays, Evan and Starace, Giulio and Liu, Kevin and Maksin, Leon and Patwardhan, Tejal and Weng, Lilian and M\k{a}dry, Aleksander},
  booktitle={The Thirteenth International Conference on Learning Representations},
  year={2025}
}

@inproceedings{mudgal2018deepmatcher,
  title={Deep Learning for Entity Matching: A Design Space Exploration},
  author={Mudgal, Sidharth and Li, Han and Rekatsinas, Theodoros and Doan, AnHai and Park, Youngchoon and Krishnan, Ganesh and Deep, Rohit and Arcaute, Esteban and Raghavendra, Vijay},
  booktitle={Proceedings of the 2018 International Conference on Management of Data},
  pages={19--34},
  year={2018},
  publisher={ACM}
}

@article{li2021ditto,
  title={Deep Entity Matching with Pre-Trained Language Models},
  author={Li, Yuliang and Li, Jinfeng and Suhara, Yoshihiko and Doan, AnHai and Tan, Wang-Chiew},
  journal={Proceedings of the VLDB Endowment},
  volume={14},
  number={1},
  pages={50--60},
  year={2021}
}

@inproceedings{karpukhin2020dpr,
  title={Dense Passage Retrieval for Open-Domain Question Answering},
  author={Karpukhin, Vladimir and O\u{g}uz, Barlas and Min, Sewon and Lewis, Patrick and Wu, Ledell and Edunov, Sergey and Chen, Danqi and Yih, Wen-tau},
  booktitle={Proceedings of the 2020 Conference on Empirical Methods in Natural Language Processing},
  pages={6769--6781},
  year={2020}
}

@article{johnson2019faiss,
  title={Billion-Scale Similarity Search with {GPUs}},
  author={Johnson, Jeff and Douze, Matthijs and J{\'e}gou, Herv{\'e}},
  journal={IEEE Transactions on Big Data},
  volume={7},
  number={3},
  pages={535--547},
  year={2021},
  publisher={IEEE}
}

@inproceedings{geifman2017selective,
  title={Selective Classification for Deep Neural Networks},
  author={Geifman, Yonatan and El-Yaniv, Ran},
  booktitle={Advances in Neural Information Processing Systems},
  volume={30},
  year={2017}
}

@article{li2025comind,
  title={{CoMind}: Towards Community-Driven Agents for Machine Learning Engineering},
  author={Li, Sijie and Sun, Weiwei and Li, Shanda and Talwalkar, Ameet and Yang, Yiming},
  journal={arXiv preprint arXiv:2506.20640},
  year={2025}
}

@inproceedings{zhao2024competeai,
  title={{CompeteAI}: Understanding the Competition Dynamics of Large Language Model-based Agents},
  author={Zhao, Qinlin and Wang, Jindong and Zhang, Yixuan and Jin, Yiqiao and Zhu, Kaijie and Chen, Hao and Xie, Xing},
  booktitle={Proceedings of the 41st International Conference on Machine Learning},
  pages={61092--61107},
  year={2024},
  publisher={PMLR}
}

@article{mouret2015mapelites,
  title={Illuminating search spaces by mapping elites},
  author={Mouret, Jean-Baptiste and Clune, Jeff},
  journal={arXiv preprint arXiv:1504.04909},
  year={2015}
}

\end{document}